\documentclass[conference,letterpaper]{IEEEtran}

\usepackage[letterpaper, left=0.628in, right=0.628in, bottom=1in, top=0.75in]{geometry}

\usepackage[T1]{fontenc}
\usepackage{graphicx}
\usepackage{cite}
\usepackage{subcaption}

\usepackage{overpic}
\usepackage{amssymb}
\usepackage{amsmath}
\usepackage{amsthm}
\usepackage{array}
\usepackage{color}
\usepackage{url}

\IEEEoverridecommandlockouts

\theoremstyle{definition}

\newtheorem{definition}{Definition}
\newtheorem{lemma}{Lemma}
\newtheorem{corollary}{Corollary}

\newtheorem{proposition}{Proposition}
\newtheorem{assumption}{Assumption}

\newcommand{\vect}[1]{\mathbf{#1}}

\def\diag{\mathrm{diag}}

\def\tr{\mathrm{tr}}

\def\Htran{\mbox{\tiny $\mathrm{H}$}}
\def\Ttran{\mbox{\tiny $\mathrm{T}$}}
\def\CN{\mathcal{N}_{\mathbb{C}}} 

\begin{document}

\title{A New Look at Cell-Free Massive MIMO: \\ Making It Practical With Dynamic Cooperation}

\author{
\IEEEauthorblockN{Emil Bj{\"o}rnson\IEEEauthorrefmark{1},
Luca Sanguinetti\IEEEauthorrefmark{2}
\thanks{E. Bj{\"o}rnson was supported by ELLIIT and the Wallenberg AI, Autonomous Systems and Software Program (WASP). L. Sanguinetti was supported by the University of Pisa under the PRA 2018-2019 Research Project CONCEPT.}}
\IEEEauthorblockA{\IEEEauthorrefmark{1}\small{Department of Electrical Engineering (ISY), Link\"{o}ping University, Link\"{o}ping, Sweden (emil.bjornson@liu.se)}}
\IEEEauthorblockA{\IEEEauthorrefmark{2}\small{Dipartimento di Ingegneria dell'Informazione, University of Pisa, 56122 Pisa, Italy (luca.sanguinetti@unipi.it)}}
}

\maketitle

\begin{abstract}
This paper takes a new look at Cell-free Massive MIMO (multiple-input multiple-output) through the lens of the \emph{dynamic cooperation cluster} framework from the Network MIMO literature. The purpose is to identify and address scalability issues that appear in prior work. We provide distributed algorithms for initial access, pilot assignment, cluster formation, precoding, and combining that are scalable in the sense of being implementable with arbitrarily many users. Interestingly, the suggested precoding and combining outperform conjugate beamforming and matched filtering, respectively, while also being fully distributed.%
\end{abstract}
\vspace{0.2cm}

\IEEEpeerreviewmaketitle

\section{Introduction}

By transmitting a signal coherently from multiple antennas, the received power can be increased without increasing the total transmit power \cite{massivemimobook}. This is the phenomenon utilized by classic beamforming from co-located antenna arrays but can be also utilized when transmitting coherently from multiple access points (APs) \cite{Shamai2001a}. Even if the APs have different channel gains to the receiver, the benefit of coherent transmission makes it better to spread out the transmit power over multiple APs than transmitting only from the AP with the best channel \cite{Ngo2017b}. Such coherent joint transmission from multiple APs has many different names, including Network MIMO  \cite{Venkatesan2007a}. 

The early Network MIMO papers assumed all APs have network-wide channel state information (CSI) and transmit to all user equipments (UEs). These are two preferable but impractical/unscalable assumptions that lead to immense backhaul signaling for CSI and data sharing, respectively.
Fortunately, \cite{Bjornson2010c} proved that Network MIMO can operate without CSI sharing, by sacrificing the ability for APs to jointly cancel interference.
Moreover, to limit data sharing, each UE can be served only by a subset of the APs. Initially, a \emph{network-centric} approach was taken by dividing the APs into non-overlapping cooperation clusters in which the APs are sharing data to serve only UEs residing in the joint coverage area.
  This approach was considered in LTE but provides small gains in practice \cite{Fantini2016a}, partially due to substantial interference between clusters. The alternative is \emph{dynamic cooperation clusters} (DCC) \cite{Bjornson2011a}, which is a \emph{user-centric} approach where each UE is served by the AP subset providing the best channel conditions. DCC didn't gain much attention at the time, since Massive MIMO (mMIMO) was simultaneously proposed  \cite{Marzetta2010a} and rightfully gained the spotlight, but it was implemented in the pCell technology \cite{Perlman2015a}.

Now that mMIMO is a rather mature technology \cite{massivemimobook}, the research focus is shifting back to Network MIMO, but under the new name of \emph{Cell-free mMIMO} \cite{Ngo2017b,Nayebi2017a}. The key novelty is the rigorous ergodic spectral efficiency (SE) analysis with imperfect CSI, but conceptually, it is a special case of Network MIMO. In fact, it was initially a step backward in terms of implementation feasibility since all APs were assumed to serve all UEs and emphasis was put on developing network-wide power control algorithms \cite{Ngo2017b,Nayebi2017a,Nayebi2016a,Bashar2019a}. The user-centric approach was reintroduced for Cell-free mMIMO in \cite{Buzzi2017a} but without making connections to DCC or other  implementation-related aspects that had already been considered in the Network MIMO literature and summarized in the textbook \cite{Bjornson2013d}.

\textbf{Contributions:} In this paper, we first expose the potential scalability issues of Cell-free mMIMO and then prove that Cell-free mMIMO is a special case of the DCC framework in \cite{Bjornson2011a,Bjornson2013d}. We utilize this perspective to present new distributed and scalable algorithms for initial access, pilot assignment, and cooperation cluster formation. We derive downlink and uplink SEs with multi-antenna APs and propose new scalable forms of signal-to-leakage-and-noise ratio (SLNR) precoding and regularized zero-forcing (RZF) combining. These methods are fully distributed and, importantly, outperform the standard conjugate beamforming and matched filtering methods.

\section{System Model and Scalability}

\begin{figure}[t!]
	\centering 
	\begin{overpic}[width=.9\columnwidth,tics=10]{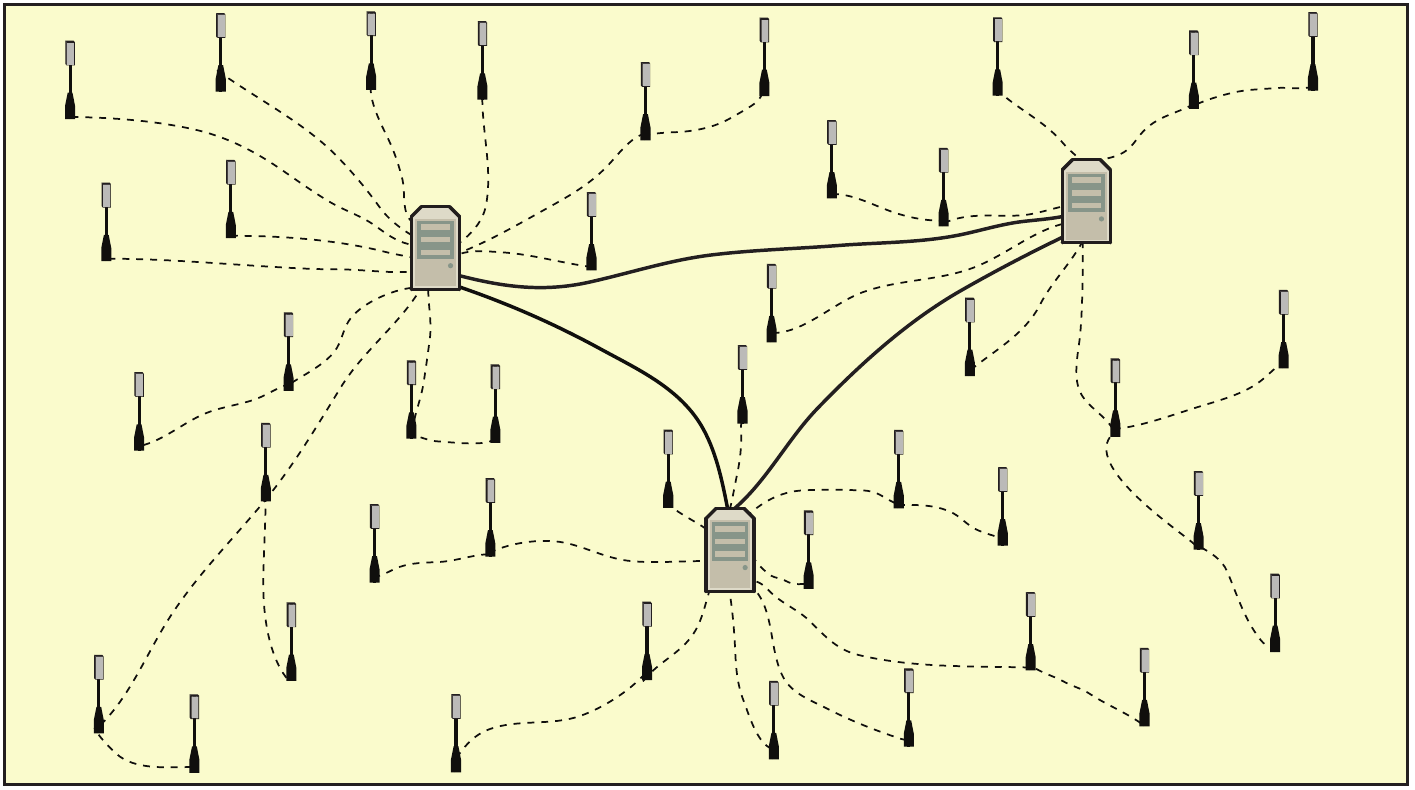}
	\put(79,41){Edge-cloud}
	\put(79,37){processor}
	\put(35,51){AP $l$}
\end{overpic} \vspace{-1mm}
	\caption{Illustration of a Cell-free mMIMO network with many distributed APs connected to edge-cloud processors. The APs are jointly serving all the UEs in the coverage area.} \vspace{-2mm}
	\label{fig:cell-free} 
\end{figure}

We consider a cell-free network consisting of $K$ single-antenna UEs and $L$ APs, each equipped with $N$ antennas. The APs are connected to edge-cloud processors \cite{Perlman2015a,Burr2018a,Interdonato2019a}, as illustrated in Fig.~\ref{fig:cell-free}, which enables coherent joint transmission and reception to the UEs in the entire coverage area.

The channel between AP $l$ and UE $k$ is denoted $\vect{h}_{kl} \in \mathbb{C}^N$ and the collective channel from all APs is $\vect{h}_k = [\vect{h}_{k1}^{\Ttran} \, \ldots \, \vect{h}_{kL}^{\Ttran}]^{\Ttran} \in \mathbb{C}^{M}$, where $M = NL$. The network operates according to a time-division duplex (TDD) 
protocol with a data transmission phase and a pilot phase for channel estimation. We consider the standard TDD protocol  \cite{massivemimobook} in which each coherence block is divided into $\tau_p$ channel uses for uplink pilots, $\tau_u$ for uplink data, and $\tau_d$ for downlink data with $\tau_c = \tau_p + \tau_u + \tau_d$.
In each block, an independent flat-fading realization is drawn using correlated Rayleigh fading:
\begin{equation}
\vect{h}_{kl} \sim \CN(\vect{0}, \vect{R}_{kl})
\end{equation}
where the spatial  correlation matrix $\vect{R}_{kl} \in \mathbb{C}^{N \times N}$  describes geometric attenuation, shadowing, and spatial properties.

\vspace{-0.5mm}

\subsection{Original Cell-free mMIMO Model}

To motivate the approach taken in this paper, we first review the system model in the original papers on Cell-free mMIMO \cite{Ngo2017b,Nayebi2017a}, where network-wide downlink transmission from all APs to all the UEs is considered.
Let $\vect{w}_{il} \in \mathbb{C}^N$ denote the precoding vector that AP $l$ assigns to UE $k$, then the received downlink signal at UE $k$ is
\begin{align}
y_k^{\rm{dl}}& = \sum_{l=1}^{L}  \sum_{i=1}^{K} \vect{h}_{kl}^{\Ttran} \vect{w}_{il} \varsigma_i + n_k  
=  \sum_{i=1}^{K} \vect{h}_k^{\Ttran} \vect{w}_i \varsigma_i + n_k \label{eq:Cell-free}
\end{align}
where $\varsigma_i \in \mathbb{C}$ is the independent unit-power data signal intended for UE $i$, $\vect{w}_k = [\vect{w}_{k1}^{\Ttran} \, \ldots \, \vect{w}_{kL}^{\Ttran}]^{\Ttran} \in \mathbb{C}^{M}$ is the collective precoding vector, and $n_k \sim \CN(0,\sigma^2)$ is the receiver noise.

The collective channel is distributed as $\vect{h}_k \sim \CN(\vect{0}, \vect{R}_{k})$ where $\vect{R}_{k} = \diag(\vect{R}_{k1}, \ldots,  \vect{R}_{kL}) \in \mathbb{C}^{M \times M}$ is the block-diagonal spatial correlation matrix. The system model \eqref{eq:Cell-free} is equivalent to a single-cell downlink mMIMO system with correlated fading. The achievable SEs in Cell-free mMIMO, thus, follow easily from the literature on mMIMO with correlated fading, recently summarized in \cite{massivemimobook}. The key difference from that literature is which precoding vectors can be selected, since these should satisfy per-AP power constraints and (preferably) use only local CSI. Network-wide downlink power optimization methods were developed in \cite{Ngo2017b,Nayebi2017a}, among others.

Similarly, during uplink data transmission, the received signal ${\bf y}_{l}^{\rm{ul}} \in \mathbb {C}^{N}$ at AP $l$ is
\begin{align}
	{\bf y}_{l}^{\rm{ul}} = \sum\limits_{i=1}^{K} {\bf h}_{il}s_{i} + {\bf n}_{l}\label{y_{l}^{ul}}
\end{align}
where $s_{i} \in \mathbb{C}$ is the signal transmitted from UE $i$ with power $p_i$ and ${\bf n}_{l}  \sim \mathcal {CN}({\bf 0},\sigma^2{\bf I}_{N})$. Network-wide uplink decoding was considered in the original papers on Cell-free mMIMO  \cite{Ngo2017b,Nayebi2016a}. In that case, AP $l$ selects a receive combining vector $\vect{v}_{kl}$ for UE $k$ and computes $\vect{v}_{kl}^{\Htran}{\bf y}_{l}^{\rm{ul}}$ locally. The network then estimates $s_k$ by computing the summation
\begin{align}\label{eq:Cell-free_UL}
\hat{s}_k = \sum_{l=1}^{L}  
\vect{v}_{kl}^{\Htran}{\bf y}_{l}^{\rm{ul}}. 
\end{align}
Note that \eqref{eq:Cell-free_UL} is equivalent to an uplink single-cell mMIMO system model with correlated fading, thus the achievable SEs easily follow from that literature \cite{massivemimobook}. The difference is which combining vectors can be used, since these should (preferably) use only local CSI. Network-wide uplink power optimization methods were developed in \cite{Ngo2017b,Nayebi2016a,Bashar2019a}, among others.

\subsection{Scalability Issues}

Although the network-wide processing in the original Cell-free mMIMO papers is appealing, it is not practical for large-scale network deployments with many UEs. To determine if the processing is scalable or not, it is helpful to let $K \to \infty$ and see which of the following operations are implementable.
\begin{enumerate}
\item Precoding and combining: AP $l$ computes $K$ precoding vectors ($\vect{w}_{lk}$ for all $k$) and $K$ combining vectors ($\vect{v}_{lk}$ for all $k$). The complexity becomes infinite as $K \to \infty$.
\item Estimation: AP $l$ must compute channel estimates for all $K$ UEs, with infinite complexity as $K \to \infty$.
\item Fronthaul signaling: AP $l$ needs to receive $K$ downlink data signals over the fronthaul network and forward $K$ received signals $\vect{v}_{kl}^{\Htran}{\bf y}_{l}^{\rm{ul}}$ over the fronthaul network.
\item Power optimization: Any network-wide power optimization has a complexity that goes to infinity as $K \to \infty$.
\end{enumerate}
The original form of Cell-free mMIMO is clearly not scalable.

\begin{definition} \label{def:scalable}
A Cell-free mMIMO network is said to be \emph{scalable} if none of four above-listed issues  appears.
\end{definition}

In the remainder of this paper, we outline a scalable implementation framework according to Definition~\ref{def:scalable}. We start from the DCC framework for Network MIMO in \cite{Bjornson2011a,Bjornson2013d}, which was claimed to be scalable but we fill in many missing details.

\subsection{Dynamic Cooperation Clusters}

The DCC framework was proposed in \cite{Bjornson2011a,Bjornson2013d} to enable ``\emph{unified analysis of anything from interference channels to ideal network MIMO}''. To this end, the diagonal matrix $\vect{D}_{il} \in \mathbb{C}^{N \times N}$ was defined, where the $j$th diagonal element is 1 if the $j$th antenna of  AP $l$ is allowed to transmit to and decode signals from UE $i$ and 0 otherwise.  By modifying \eqref{eq:Cell-free}, the received downlink signal at UE $k$ becomes
\begin{equation} \label{eq:DCC}
y_k^{\rm{dl}} = \sum_{l=1}^{L}  \sum_{i=1}^{K} \vect{h}_{kl}^{\Ttran} \vect{D}_{il}  \vect{w}_{il} \varsigma_i + n_k = \sum_{i=1}^{K} \vect{h}_k^{\Ttran} \vect{D}_i \vect{w}_i \varsigma_i + n_k
\end{equation}
where $\vect{D}_i = \diag(\vect{D}_{i1}, \ldots, \vect{D}_{iL}) \in \mathbb{C}^{M \times M}$ is block-diagonal. By selecting $\vect{D}_1,\ldots,\vect{D}_K$ in different ways,  \eqref{eq:DCC} can be used to model many different types of multi-AP networks; see \cite{Bjornson2013d}.

The original Cell-free mMIMO in \eqref{eq:Cell-free} is obtained from \eqref{eq:DCC} in the special case of $\vect{D}_i=\vect{I}_M \, \forall i$, where all antennas serve all UEs. The user-centric approach to Cell-free mMIMO described in \cite{Buzzi2017a} is also an instance of the DCC framework. In \cite{Buzzi2017a}, $\mathcal{M}(k) \subset \{ 1, \ldots, L\}$ denotes the subset of APs that communicate with UE $k$, which corresponds to setting
\begin{equation} \label{eq:user-centric}
\vect{D}_{kl}= \begin{cases}
\vect{I}_N & \textrm{if } l \in \mathcal{M}(k), \\ 
\vect{0}_N & \textrm{if } l \not \in \mathcal{M}(k). \end{cases}
 \end{equation}
This is exactly the same setup as considered in \cite{Bjornson2011a}.

The DCC framework does not change the received uplink signal in $\eqref{y_{l}^{ul}}$, but the uplink data estimate  in \eqref{eq:Cell-free_UL} changes to 
\begin{align} \label{eq:uplink-data-estimate}
\hat{s}_k &= \sum_{l=1}^{L}
\vect{v}_{kl}^{\Htran} \vect{D}_{kl} {\bf y}_{l}^{\rm{ul}} = \sum_{l \in \mathcal{M}(k)} 
\vect{v}_{kl}^{\Htran}{\bf y}_{l}^{\rm{ul}}
\end{align}
where the second equality only holds when using \eqref{eq:user-centric}.

\begin{figure}[t!]
	\centering 
	\begin{overpic}[width=\columnwidth,tics=10]{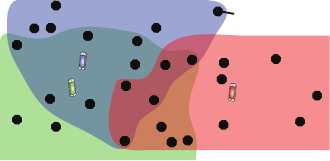}
	\put(72,43){AP $l$}
	\put(27,29){UE 1}
	\put(20,45){AP cluster for UE 1}
	\put(11,22){UE 2}
	\put(1,1){AP cluster for UE 2}
	\put(73,19){UE 3}
	\put(64,4){AP cluster for UE 3}
\end{overpic} 
	\caption{Example of dynamic cooperation clusters for three UEs.}
	\label{fig:illustrateCooperation}  \vspace{-2mm}
\end{figure}
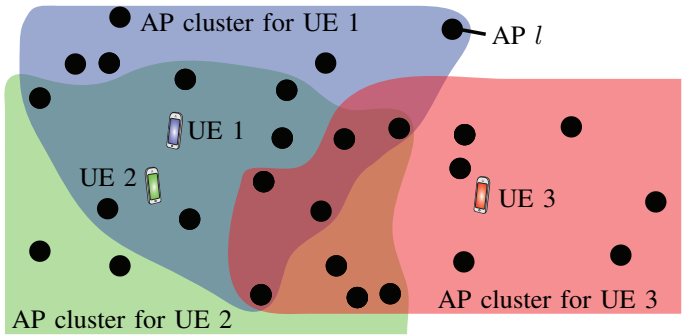

Fig.~\ref{fig:illustrateCooperation} illustrates a network with three UEs that are served by a large number of APs. The colored regions illustrate which clusters of APs are transmitting to which UEs. The fact that the clusters are partially overlapping is a core feature of DCCs, and also demonstrates that this is a cell-free network.

The DCC framework was proposed in \cite{Bjornson2011a} to achieve scalability in Network MIMO, for example, in the following way.

\begin{lemma} \label{lemma:D-requirement}
The UEs served by AP $l$ have indices in the set
\begin{equation}
\mathcal{D}_l = \left\{ i : \tr(\vect{D}_{il})\geq 1, i \in \{ 1, \ldots, K \} \right\}.
\end{equation}
If the cardinality $|\mathcal{D}_l |$ is constant as $K \to \infty$, the precoding/combining complexity and fronthaul signaling parts of Definition~\ref{def:scalable} are satisfied and thus scalable.
\end{lemma}
\begin{IEEEproof}
AP $l$ only needs to compute precoding and combining vectors for $|\mathcal{D}_l |$ UEs, and it only needs to send/receive data related to these UEs over the fronthaul network.
\end{IEEEproof}

With this result in mind, the practically important question is how to select the sets $\{\mathcal{D}_l:\forall l \}$ in a scalable way, while guaranteeing service to all UEs. This challenge is tackled below.

\section{Distributed and Scalable Implementation}

In this section, we propose a scalable, distributed implementation of Cell-free mMIMO. It is inspired by the guidelines for distributed Network MIMO in \cite[Sec.~4.3, 4.7]{Bjornson2013d}, but is far more  detailed and also focused on resource allocation.
We begin by making the following key assumption.

\begin{assumption} \label{assumption1}
Each AP serves at most one UE per pilot and uses all its antennas to serve these UEs. This implies $|\mathcal{D}_l |  \leq \tau_p$ and $\vect{D}_{il}=\vect{I}_N$ for all $i \in \mathcal{D}_l$,  $l=1,\ldots,L$. Hence, the scalability requirement in Lemma~\ref{lemma:D-requirement} is satisfied.
\end{assumption}

The rationale for Assumption~\ref{assumption1} is: a) pilot contamination makes the channel estimates of pilot-sharing UEs similar (or identical) so the AP will cause strong interference if it transmits to more than one such UE; b) the signal processing complexity becomes fixed and scalable, although all $N$ antennas are used; c) the fronthaul capacity can be dimensioned to support $\tau_p$ parallel uplink/downlink data signals per AP.

\subsection{Pilot Transmission and Channel Estimation}

There are $\tau_p$ mutually orthogonal  $\tau_p$-length pilot signals that are assigned to the UEs. Note that the pilot transmission protocol is scalable since  $\tau_p$ is a constant, independent of $K$. An algorithm for pilot assignment is provided in Section~\ref{eq:pilot-assignment}, but for now we let $\mathcal{S}_t \subset \{ 1, \ldots, K\}$ denote the subset of UEs assigned to pilot $t$. When these UEs transmit their pilot, the received pilot signal $\vect{y}_{tl}^p \in \mathbb{C}^N$ at AP $l$ is 
\begin{equation}
\vect{y}_{tl}^p = \sum_{i \in \mathcal{S}_t} \sqrt{ \tau_p p_i} \vect{h}_{il} + \vect{n}_{tl}
\end{equation}
where $p_i$ is the transmit power, $\tau_p$ is the processing gain, and $\vect{n}_{tl} \sim \CN (\vect{0}, \sigma^2 \vect{I}_N)$ is noise. Using standard results \cite[Sec.~3]{massivemimobook}, the minimum mean-squared error (MMSE) estimate of $\vect{h}_{kl}$ for $k \in \mathcal{S}_t$ is  
\begin{equation} \label{eq:estimates}
\hat{\vect{h}}_{kl} = \sqrt{p_k \tau_p} \vect{R}_{kl} \vect{\Psi}_{tl}^{-1} \vect{y}_{tl}^p  \sim \CN \left( \vect{0}, p_k \tau_p \vect{R}_{kl} \vect{\Psi}_{tl}^{-1} \vect{R}_{kl} \right)
\end{equation}
where the correlation matrix $\vect{\Psi}_{tl} = \mathbb{E} \{ \vect{y}_{tl}^p (\vect{y}_{tl}^p)^{\Htran} \}$ is given by
\begin{equation} \label{eq:Psitl}
\vect{\Psi}_{tl} = \sum_{i \in \mathcal{S}_t} \tau_p p_i \vect{R}_{il} + \sigma^2 \vect{I}_{N}.
\end{equation}
The AP uses these estimates for receiving the uplink data and for precoding the downlink data. AP $l$ only needs to compute estimates $\hat{\vect{h}}_{kl}$ for $k \in \mathcal{D}_l $, which under Assumption~\ref{assumption1} is at most one UE per pilot. Since the complexity per AP is independent of $K$, the pilot transmission is scalable when $K \to \infty$.

\subsection{Initial Access and Pilot Assignment} \label{eq:pilot-assignment}

When a new UE wants to access the network, it needs to be assigned a pilot and make it into the set $\mathcal{D}_l $ of at least one AP. This must be done in a distributed fashion, which has the risk that the UE is inadvertently dropped from service since no AP decides to transmit to it. To avoid that, each UE appoints a \emph{Master AP} that is required to transmit to it and coordinate the decoding of the uplink data  \cite{Bjornson2013d}.  Let $K+1$ be the index of the connecting UE, then the proposed access procedure is:
\begin{enumerate}

\item The UE measures $\beta_{l} = \tr(\vect{R}_{(K+1)l})/N$ for all nearby APs, using periodically broadcasted synchronization signals, and appoints AP $\ell = \mathrm{arg \, max}_{l} \, \beta_{l}$ as its Master AP. The UE also uses this signal to synchronize to the AP.

\item The UE contacts its Master AP via a standard random access procedure. The AP responds by assigning pilot  $\tau = \mathrm{arg \, min}_{t} \, \tr( \vect{\Psi}_{tl} )$ to the UE, with $\vect{\Psi}_{tl}$ given in \eqref{eq:Psitl}.

\item The Master AP informs a limited set of neighboring APs that it is now serving UE $K+1$ on pilot $\tau$. These APs independently decide if they will also serve the UE.

\end{enumerate}
In summary, the UE appoints the AP with the strongest channel as its Master AP and it is assigned to the pilot that this AP observes the least pilot power on.\footnote{This should be a pilot on which the AP is not currently serving a UE as being its Master AP, since that role has higher priority. Each AP can only be the Master AP of up to $\tau_p$ UEs in the proposed framework, but in the unlikely event that this cannot be satisfied, multiple UEs can be assigned to the same pilot but multiplexed in time and/or frequency instead. Alternatively, the second strongest AP can be appointed the Master AP.} When other APs decide whether to also serve the new UE, Assumption~\ref{assumption1} must be enforced. To limit interference (and pilot contamination), it is reasonable for an AP to switch to serving the new UE if it has a better channel to it than to the UE it currently serves on that pilot \emph{and} it is not the Master AP of the current UE. When a UE moves around, the proposed access procedure can be redone when needed; the UE then acts as if it is connecting and appoints a new Master AP, which might assign a new pilot. The old Master AP transfers its status to the new Master AP.

\subsection{Downlink SE and Distributed Precoding}

Next, we derive and analyze a general achievable downlink SE expression (i.e., a lower bound on the ergodic capacity) for the DCC system model in \eqref{eq:DCC}. AP $l$ selects its precoding vectors $ \vect{w}_{kl}$ for $k \in \mathcal{D}_l $ as a function of $\{\hat{\vect{h}}_{kl} : k \in \mathcal{D}_l \}$, while $\vect{D}_{il} \vect{w}_{il} = \vect{0}$ for $i \not \in \mathcal{D}_l$ in all expressions so these precoding vectors need not be selected. The precoding vectors of the UEs that the AP serves must also satisfy the power constraint
\begin{equation}
\sum_{k \in \mathcal{D}_l} \mathbb{E} \{ \| \vect{w}_{kl}  \|^2 \} \leq \rho
\end{equation}
where $\rho$ is the total transmit power of an AP.

We use the \emph{hardening bound} that is widely used in the mMIMO literature to compute SEs \cite[Th.~4.6]{massivemimobook}; it has also been used in \cite{Ngo2017b,Nayebi2017a} for Cell-free mMIMO with $N=1$, $\vect{D}_i=\vect{I}_M \, \forall i$, for specific choices of precoding schemes.
\begin{proposition} 
An achievable downlink SE $R_k^{\textrm{dl}}$ [bit/s/Hz] for UE $k$ is 
\begin{equation} \label{eq:uatf-dl}
\frac{\tau_d}{\tau_c} \log_2 \left( 1+ \frac{ | \mathbb{E} \{ \vect{h}_k^{\Ttran} \vect{D}_k \vect{w}_k \} |^2 }{ \sum\limits_{i=1}^{K} \mathbb{E} \{ |\vect{h}_k^{\Ttran} \vect{D}_i \vect{w}_i|^2 \} -  | \mathbb{E} \{ \vect{h}_k^{\Ttran} \vect{D}_k \vect{w}_k \} |^2 + \sigma^2 }  \right).
\end{equation}
\end{proposition}
\begin{IEEEproof}
This is proved by following the same approach as in \cite[Th.~4.6]{massivemimobook}, but for the system model in \eqref{eq:DCC}.
\end{IEEEproof}

The expectations in \eqref{eq:uatf-dl} can be computed by Monte-Carlo simulations for any choice of precoding vectors. The precoding at AP $l$ should only depend on $\{\hat{\vect{h}}_{il} : i \in \mathcal{D}_l \}$ to achieve a scalable implementation \cite{Bjornson2010c}. Without loss of generality, we set
\begin{equation}
\vect{w}_{il} = \sqrt{ \frac{ \rho_{il} }{\mathbb{E} \{ \| \bar{\vect{w}}_{il} \|^2 \}} } \bar{\vect{w}}_{il} \quad \forall i \in \mathcal{D}_l
\end{equation}
where $\rho_{il}\!\geq\!0$ is the transmit power and $\bar{\vect{w}}_{il} \!\in\!\mathbb{C}^{N}$ gives the precoding direction. Two schemes that satisfy the scalability requirement are maximum ratio (MR) and SLNR~\cite{Bjornson2010c}:
\begin{equation} \label{eq:precoding-schemes}
\bar{\vect{w}}_{kl} = \begin{cases}
\hat{\vect{h}}_{kl}^* & \textrm{with MR}, \\
\bigg(\sum\limits_{i \in \mathcal{D}_l} \rho_{il} \hat{\vect{h}}_{il}^* \hat{\vect{h}}_{il}^{\Ttran} + \sigma^2 \vect{I}_N \bigg)^{\!-1}\hat{\vect{h}}_{kl}^* & \textrm{with SLNR},
\end{cases}
\end{equation}
for $k \in \mathcal{D}_l$. MR is also known as conjugate beamforming and is the standard scheme in the Cell-free mMIMO literature. The scalable SLNR precoding in \eqref{eq:precoding-schemes} is new since previous expressions consider all UEs in the network \cite{Bjornson2013d}.
The benefit of SLNR over MR is two-fold: 1) it suppresses interference spatially if $N>1$ since $\bar{\vect{w}}_{kl}$ maximizes the ratio between desired signal power and interference caused to the other UEs served by the AP; and 2) it reduces variations in the effective  gains $\vect{h}_{kl}^{\Ttran}\vect{w}_{il}$ of desired and interfering channels for any $N$.

\begin{figure}[t!]
	\centering 
	\begin{overpic}[width=\columnwidth,tics=10]{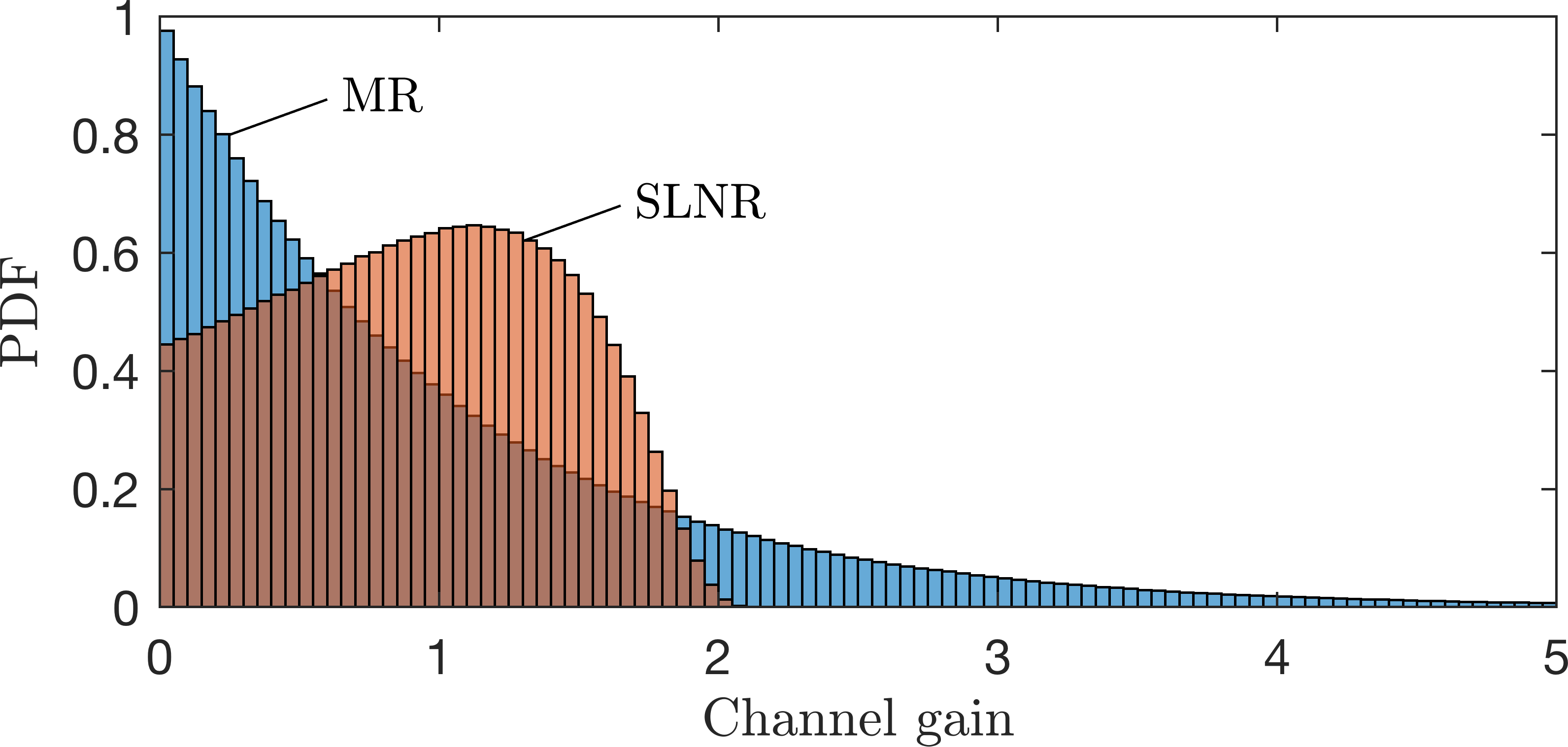}
\end{overpic} \vspace{-4mm}
	\caption{For $N=M=K=\rho=\sigma^2=1$ and perfect CSI, the channel gain is $|h|^2$ with MR and $\frac{|h|^2}{(|h|^2+1)^2} / \mathbb{E} \{ \frac{|h|^2}{(|h|^2+1)^2}  \}$ with SLNR, where $x \sim \CN(0,1)$. Their PDFs are widely different, particularly only SLNR has bounded support.}
	\label{fig:simulationGain}  \vspace{-3mm}
\end{figure}

The latter is a non-trivial phenomenon that appears even with $N=M=K=1$ and perfect CSI. Fig.~\ref{fig:simulationGain} shows the probability density function (PDF) of the  channel gains with MR and SLNR in that case. The channel gains have identical mean values, but MR gives an exponential distribution with an infinite tail while SLNR has small and compact support. This behavior will lead to higher SE when using SLNR.

The ``only'' benefit of MR is that the SE can be computed in closed form, following the same approach as in \cite[Cor.~4.7]{massivemimobook}.
\begin{corollary} \label{cor:closed-form}
With MR, the expectations in \eqref{eq:uatf-dl} become
\begin{equation} 
\mathbb{E} \{ \vect{h}_k^{\Ttran} \vect{D}_k \vect{w}_k \} = \sum_{l \in \mathcal{D}_k} \sqrt{\rho_{il} p_k \tau_p \tr( \vect{D}_{kl} \vect{R}_{kl} \vect{\Psi}_{t_k l}^{-1} \vect{R}_{kl})}
\end{equation}
\begin{align} \notag 
&\mathbb{E} \{ |\vect{h}_k^{\Ttran} \vect{D}_i \vect{w}_i|^2 \}  = \sum_{l=1}^{L} \rho_{il} \frac{  \tr \left(  \vect{D}_{il} \vect{R}_{il} \vect{\Psi}_{t_i l}^{-1} \vect{R}_{il} \vect{D}_{il} \vect{R}_{kl}  \right) }{ \tr( \vect{R}_{il} \vect{\Psi}_{t_i l}^{-1} \vect{R}_{il})} \\
&+ \begin{cases}
\left|  \sum\limits_{l=1}^{L} \sqrt{ \rho_{il} p_k \tau_p} \frac{  \tr \left(  \vect{D}_{il} \vect{R}_{il} \vect{\Psi}_{t_i l}^{-1} \vect{R}_{kl}  \right) }{ \sqrt{ \tr( \vect{R}_{il} \vect{\Psi}_{t_i l}^{-1} \vect{R}_{il}) } } \right|^2 &  \text{if }  t_i = t_k
\\
0 & \textrm{otherwise},
\end{cases}
\end{align}
where $t_i$ is the index of the pilot assigned to UE $i$.
\end{corollary}

\begin{figure*}
\begin{align} \label{eq:uatf-ul} \tag{20}
R_k^{\textrm{ul}} = \frac{\tau_u}{\tau_c} \log_2 \left( 1+ \frac{ p_k \bigg| \sum\limits_{l =1}^{L}  \mathbb{E} \left\{  \vect{v}_{kl}^{\Htran} \vect{D}_{kl} \vect{h}_{kl} \right\} \bigg|^2 }{ \sum\limits_{i=1}^{K} p_i \mathbb{E} \bigg\{ \bigg| \sum\limits_{l =1}^{L}  \vect{v}_{kl}^{\Htran} \vect{D}_{kl} \vect{h}_{il} \bigg|^2 \bigg\} -  p_k \bigg| \sum\limits_{l =1}^{L}  \mathbb{E} \left\{  \vect{v}_{kl}^{\Htran} \vect{D}_{kl} \vect{h}_{kl} \right\} \bigg|^2 + \sigma^2 \sum\limits_{l =1}^{L}  \mathbb{E} \left\{ \left\|  \vect{D}_{kl}^{\Htran} \vect{v}_{kl} \right\|^2 
\right\} }  \right)
\end{align} \vspace{-1mm}
\hrule \vspace{-3mm}
\end{figure*}

AP $l$ needs to select the transmit powers $\rho_{il}$ for $i \in \mathcal{D}_l$. 
Network-wide optimization algorithms, considered in \cite{Bjornson2013d,Ngo2017b,Nayebi2017a}, are not scalable as $K \to \infty$ since the number of optimization variables grows with $K$.\footnote{It is possible to implement network-wide optimization problems in an iterative semi-distributed way, for example, using dual decomposition theory \cite[Sec.~4.3]{Bjornson2013d}. However, these approaches converge slowly, require even more optimization variables, and need a lot of backhaul signaling. Hence, this approach is neither practical nor scalable.} Since each AP is (at least partially) unaware of the power allocation decisions made at other APs, only heuristic solutions are scalable. There are plenty of such schemes in the literature; some examples
 are found in \cite{Bjornson2010c,Bjornson2011a,Nayebi2017a,Interdonato2019a}, \cite[Sec.~3.4.4]{Bjornson2013d}. Since evaluation and comparison of heuristic schemes require extensive simulations, which is outside the scope of this paper, we consider only equal power allocation at each AP:
\begin{equation} \label{eq:equal_power_DL}
\rho_{il} = \begin{cases} \frac{\rho}{| \mathcal{D}_l |} & \textrm{if } i \in \mathcal{D}_l, \\
0 &  \textrm{otherwise}.
\end{cases}
\end{equation}
Since each UE is guaranteed to be served by at least one AP (i.e., its Master AP), each UE will be assigned non-zero transmit power when using \eqref{eq:equal_power_DL} and, thus, get a non-zero SE.

\subsection{Uplink SE and Distributed Combining}

Next, we derive and analyze an achievable uplink SE expression based on the combined uplink signal in \eqref{eq:uplink-data-estimate} from the serving APs. We use the \emph{use-and-then-forget bound} that is widely used in the mMIMO literature \cite[Th.~4.4]{massivemimobook}, and also used in \cite{Nayebi2016a,Bashar2019a} for Cell-free mMIMO with $N=1$, $\vect{D}_i=\vect{I}_M \, \forall i$, for specific choices of combining schemes.

\setcounter{equation}{20}

\begin{proposition} 
An achievable uplink SE $R_k^{\textrm{ul}}$ [bit/s/Hz] for UE $k$ is given in \eqref{eq:uatf-ul} on the top of this page.
\end{proposition}
\begin{IEEEproof}
This is proved by following the same approach as in \cite[Th.~4.4]{massivemimobook}, but for the received signal in \eqref{eq:uplink-data-estimate}.
\end{IEEEproof}

A key difference between the uplink and downlink is that, in the uplink, only the APs that serve the UE affects its SE. 

AP $l$ need to select its combining vectors $ \vect{v}_{il}$ for $i \in \mathcal{D}_l $ as a function of $\{\hat{\vect{h}}_{il} : i \in \mathcal{D}_l \}$. Two schemes that satisfy this requirement are MR (i.e., matched filtering) and RZF:
\begin{equation}
\vect{v}_{kl} = \begin{cases}
\hat{\vect{h}}_{kl} & \textrm{with MR}, \\
p_k \Big(\sum\limits_{i \in \mathcal{D}_l} p_{i} \hat{\vect{h}}_{il} \hat{\vect{h}}_{il}^{\Htran} + \sigma^2 \vect{I}_N \Big)^{\!-1}\hat{\vect{h}}_{kl} & \textrm{with RZF},
\end{cases}
\end{equation}
for $k \in \mathcal{D}_l$. This new variant of RZF only includes the channel estimates of the UEs that the AP is serving, thus making it a novel contribution and different from the non-scalable L-MMSE scheme recently proposed in \cite{Bjornson2019c}. The expectations in \eqref{eq:uatf-ul} can be computed by Monte Carlo simulations for RZF, while a closed form expression similar to Corollary~\ref{cor:closed-form} can be obtained for MR, but we omit this part for space limitations.

The uplink transmit powers $\{ p_k : \forall k\}$ need to be selected and we assume that each UE has a maximum power of $P$. The network-wide power optimization methods in \cite{Ngo2017b,Nayebi2016a,Bashar2019a} are not scalable, thus a heuristic solution is needed. 
Since transmission at full power provided good performance for both strong and weak UEs in \cite{Bjornson2019c}, we use that power control policy:
\begin{equation}
p_k = P, \quad k = 1,\ldots, K.
\end{equation}

\subsection{Network Topology, Signal Encoding and Decoding}

The proposed algorithms are transparent to the network topology since only neighboring APs cooperate. One option is to have local processing at each AP, as in classic cellular networks, and backhaul connections to the core network. Another option is to divide the APs into disjunct sets and connect each one via fronthaul to an edge-cloud processor \cite{Perlman2015a,Burr2018a,Interdonato2019a} for centralized processing, as illustrated in Fig.~\ref{fig:cell-free}. Many other cloud-RAN implementations are possible.

The most complicated part to implement is the encoding of downlink data and decoding of uplink data. We propose that the Master AP is carrying out these tasks, either locally or by delegating the task to a nearby edge-cloud processor. In the downlink, the data is shared to the neighboring APs that also transmit to the UE. In the uplink, the neighboring APs make soft estimates  $\vect{v}_{kl}^{\Htran}{\bf y}_{l}^{\rm{ul}}$ of the data, which are sent to the Master AP (or a nearby edge-cloud processor) for final decoding. 

\vspace{-1mm}

\section{Numerical Results}
\label{sec:numerical}

In this section, we compare the SEs achieved with standard MR and the proposed SLNR precoding and RZF combining. We consider a setup with $M=400$ APs and $K=100$ UEs  independently and uniformly distributed in a $2 \times 2$\,km square. By using the wrap-around technique, we approximate an infinitely large network with 100\,APs/km$^2$ and 25\,UEs/km$^2$. The UEs connect to the network as described in Section~\ref{eq:pilot-assignment}, starting with $\tau_p$ UEs with different pilots and then letting UEs connect one after the other. We use the propagation model from \cite[Sec.~4.1.3]{massivemimobook} with correlated fading, with the difference that the APs are deployed 10\,m above the UEs, which gives a minimum distance. We have $\tau_c = 200$, $\tau_p=10$, $\rho=p_k=100$\,mW, and 20\,MHz bandwidth. We use $\tau_d = 190$ and $\tau_u = 190$ when evaluating downlink and uplink, respectively.

\begin{figure} 
        \centering 
        \begin{subfigure}[b]{\columnwidth} \centering 
                \includegraphics[width=0.95\columnwidth]{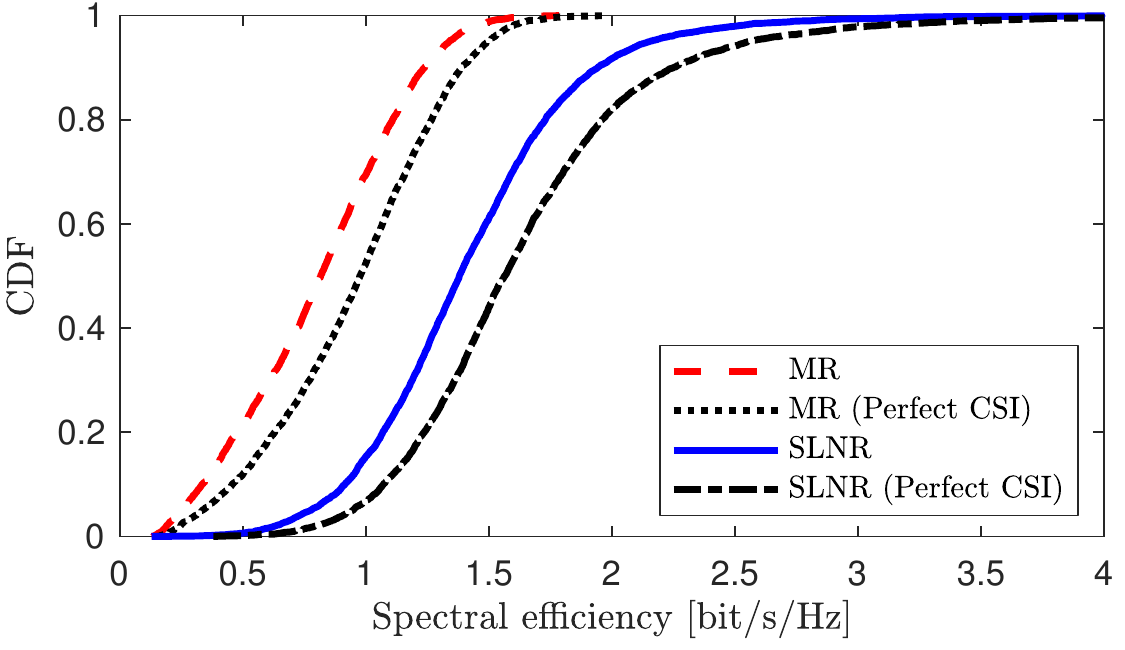}  \vspace{-1mm}
                \caption{$N=1$ antennas per AP.} \vspace{1.5mm}
                \label{fig:simulationSE_N1}
        \end{subfigure} 
        \begin{subfigure}[b]{\columnwidth} \centering
                \includegraphics[width=0.95\columnwidth]{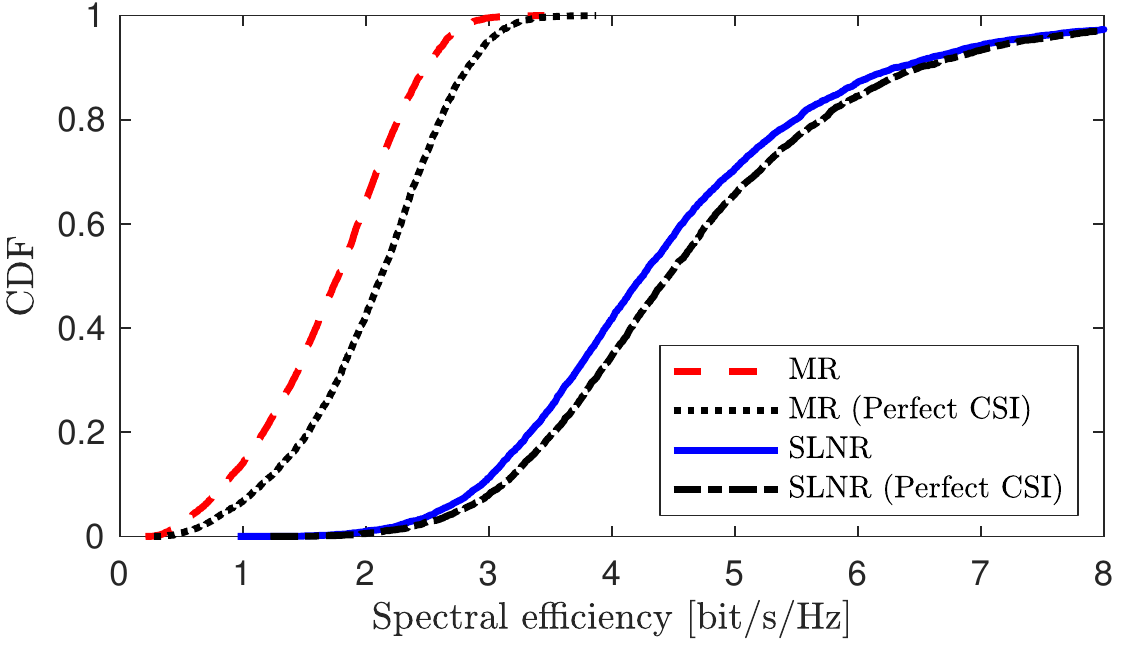} \vspace{-1mm}
                \caption{$N=4$ antennas per AP.}  
                \label{fig:simulationSE_N4} 
        \end{subfigure} 
        \caption{Downlink SE per UE for different precoding schemes.} 
        \label{fig:simulationSE}   \vspace{-3mm}
\end{figure}

Fig.~\ref{fig:simulationSE} shows the cumulative distribution function (CDF) of the downlink SE per UE in \eqref{eq:uatf-dl}, where the randomness considers different AP and UE locations. We compare MR or SLNR precoding. Fig.~\ref{fig:simulationSE}(a) considers $N=1$ and we notice that SLNR achieves 80\% higher average SE than MR; the improvement is largest for the UEs with the best channel conditions. This means that the phenomenon illustrated in Fig.~\ref{fig:simulationGain} has a huge impact on performance: MR achieves slightly higher signal power than SLNR for every UE, but also a much higher interference for most UEs due to the larger variations.

Fig.~\ref{fig:simulationSE}(b) considers $N=4$ and the multiple AP antennas improve the SE of all the UEs, since the precoding gain can increase the received signal power with up to $4\times$. The gain is particularly large with SLNR since each AP can now also suppress interference spatially. SLNR now achieves 155\% higher average SE than MR. 
The genie-aided case with perfect CSI at the UEs is shown as a reference. We notice that the SEs are close to these upper bounds, thus the difference between MR and SLNR is not an artifact of the hardening bound.

\begin{figure}[t!]
	\centering 
	\begin{overpic}[width=0.95\columnwidth,tics=10]{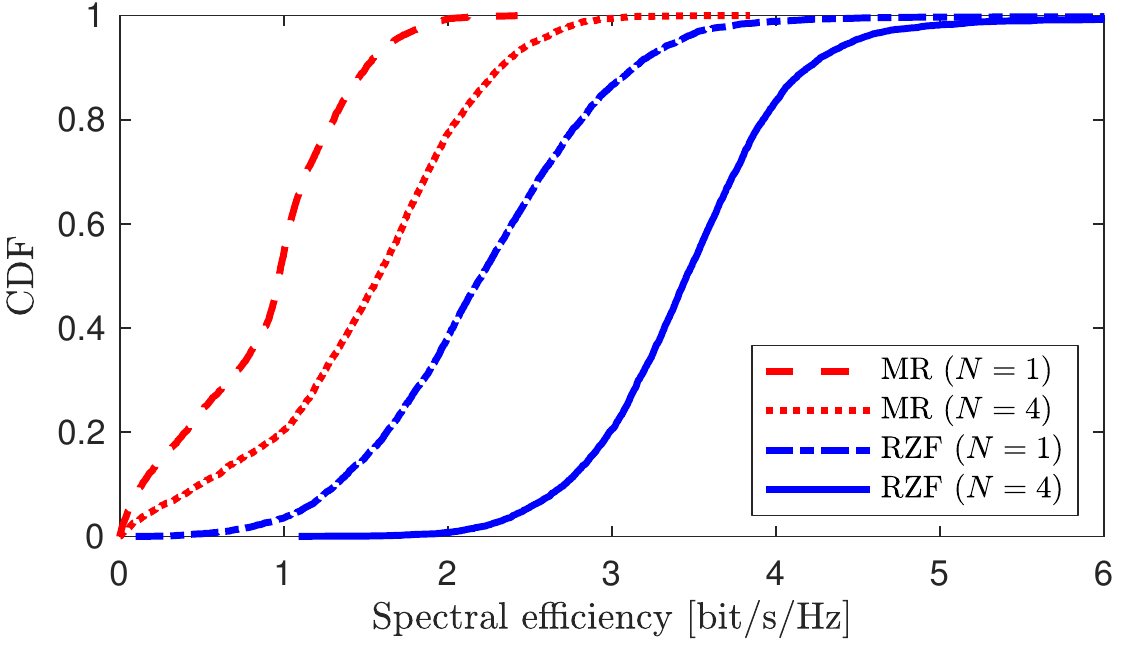}
\end{overpic} \vspace{-1mm}
	\caption{Uplink SE per UE for different combining schemes.} \vspace{-4mm}
	\label{fig:simulationSE_uplink} 
\end{figure}

Fig.~\ref{fig:simulationSE_uplink} shows the CDF of the uplink SE per UE in \eqref{eq:uatf-ul} with $N=1$ or $N=4$. We notice that RZF outperforms MR, which is expected in light of the downlink results---RZF is basically the uplink counterpart of SLNR. The difference is so large that SLNR with $N=1$ gives 48\% higher average SE than MR with $N=4$.  In addition to the results shown in Fig.~\ref{fig:simulationSE_uplink}, we compared with the SE in the genie-aided case when perfect CSI is available in the decoding. RZF gives SE close to that genie-bound, while MR does not. Hence, it is only when using MR that the lack of channel hardening in Cell-free mMIMO (which was proved in \cite{Chen2017a}) make the SE expressions loose.

\section{Conclusions}
\vspace{-0.5mm}

This paper has proposed a scalable, distributed implementation of Cell-free mMIMO by exploiting the DCC framework from the Network MIMO literature. We considered initial access, pilot assignment, cooperation cluster formation, precoding, and combining.
We have demonstrated that MR is outperformed by the proposed SLNR (RZF) in the downlink (uplink), even for single-antenna APs. We stress that the new SLNR and RZF are fully distributed schemes, in contrast to the network-wide schemes considered in \cite{Nayebi2016a,Nayebi2017a,Bjornson2019c}. Hence, SLNR and RZF are respectively the new state-of-the-art in distributed precoding and combining for Cell-free mMIMO.

\vspace{-0.8mm}

\bibliographystyle{IEEEtran}
\bibliography{IEEEabrv,refs}

\end{document}